\documentclass[12pt]{article}
\usepackage{epsfig,rotating,setspace,latexsym,amsmath,amssymb,epsf,bm,subfigure}
\usepackage{cite}

\title{{Power Control for User Cooperation}\thanks{This work was supported by NSF Grants ANI 02-05330, CCR
03-11311 and CCF 04-47613; and ARL/CTA Grant DAAD 19-01-2-0011,
and was presented in part at the IEEE International Conference on
Wireless Networks, Communications, and Mobile Computing, Maui, HI,
June 2005.}}

\author{Onur Kaya\thanks{Department of Electronics Engineering, I\c{s}\i k University, Istanbul, Turkey
\{onurkaya@isikun.edu.tr\}.}
        \quad Sennur Ulukus\thanks{Department of ECE, University of Maryland, College Park, MD 20742
        \{ulukus@umd.edu\}.}}

\newcommand{\s}{\mbox{\footnotesize sum}}

\newtheorem{proposition}{Proposition}

\newtheorem{corollary}{Corollary}

\newenvironment{Proof}[1]{\medskip\par\noindent
{\bf Proof:\,}\,#1}{{\mbox{\,$\Box$}\par}}

\newcommand{\muv}{\bm{\mu}}

\newcommand{\pv}{\mathbf{p}}

\newcommand{\gv}{\mathbf{g}}
\newcommand{\hv}{\mathbf{h}}

\newcommand{\pb}{\mbox{$\bar{p}$}}

\begin{document}
\date{}
\maketitle \vspace*{-0.5cm} \setstretch{1.735}
\begin{abstract}
For a fading Gaussian multiple access channel with user
cooperation, we obtain the optimal power allocation policies that
maximize the rates achievable by block Markov superposition
coding. The optimal policies result in a coding scheme that is
simpler than the one for a general multiple access channel with
generalized feedback. This simpler coding scheme also leads to the
possibility of formulating an otherwise non-concave optimization
problem as a concave one. Using the channel state information at
the transmitters to adapt the powers, we demonstrate significant
gains over the achievable rates for existing cooperative systems.
\end{abstract}
\vspace*{0.4cm}\noindent {\em Index terms:} Cooperative diversity,
fading, multiple access channel, power control, user cooperation.
\newpage
\section{Introduction}
\vspace{3mm} Increasing demand for higher rates in wireless
communication systems have recently triggered major research
efforts to characterize the capacities of such systems. The
wireless medium brings along its unique challenges such as fading
and multiuser interference, which make the analysis of the
communication systems more complicated. On the other hand, the
same challenging properties of such systems are what give rise to
the concepts such as diversity, over-heard information, etc.,
which can be carefully exploited to the advantage of the network
capacity.

In the early 1980s, several problems which form a basis for the
idea of user cooperation in wireless networks were solved. First,
the case of a two user multiple access channel (MAC) where both
users have access to the channel output was considered by Cover
and Leung \cite{cover_leung}, and an achievable rate region was
obtained for this channel. Willems and van der Meulen then
demonstrated \cite{vandermeulen} that the same rate region is
achievable if there is a feedback link to only one of the
tansmitters from the channel output.

The capacity region of the MAC with partially cooperating encoders
was obtained by Willems in \cite{partially_cooperating}. In this
setting, the encoders are assumed to be connected by finite
capacity communication links, which allow the cooperation. Willems
and van der Meulen also considered a limiting case of cooperation
where the encoders ``crib'' from each other, that is, they learn
each others' codewords before the next transmission
\cite{cribbing_enc}. Several scenarios regarding which encoder(s)
crib, and how much of the codewords the encoders learn, are
treated and the capacity region for each case is obtained in
\cite{cribbing_enc}. The capacity of such channels are an upper
bound to the rates achievable by cooperative schemes, since in the
case of cribbing encoders, the sharing of information comes for
free, i.e., the transmitters do not allocate any resources such as
powers, to establish common information.

An achievable rate region for a MAC with generalized feedback was
found in \cite{generalized_feedback}. This channel model is worth
special attention as far as the wireless channels are concerned,
since it models the over-heard information by the transmitters. In
particular, for a two user MAC with generalized feedback described
by $({\cal X}_1\times{\cal X}_2,P(y,y_1,y_2|x_1,x_2),{\cal
Y}\times{\cal Y}_1\times{\cal Y}_2)$, where user 1 has access to
channel output $Y_1$ and user two has access to channel output
$Y_2$, an achievable rate region is obtained by using a
superposition block Markov encoding scheme, together with backward
decoding, where the receiver waits to receive all $B$ blocks of
codewords before decoding.

Recently, Sendonaris, Erkip and Aazhang have successfully employed
the results of these rather general problems, particularly that of
generalized feedback, to a Gaussian MAC in the presence of fading,
leading to user cooperation diversity and higher rates
\cite{coop_div}. In this setting, both the receiver and the
transmitters receive noisy versions of the transmitted messages,
and slightly modifying the basic relay channel case, the
transmitters form their codewords not only based on their own
information, but also on the information they have received from
each other. It is assumed in \cite{coop_div} that channel state
information for each link is known to the corresponding receiver
on that link, and also phase of the channel state needs to be
known at the transmitters in order to obtain a coherent combining
gain. The achievable rate region is shown to improve significantly
over the capacity region of the MAC with non-cooperating transmitters,
especially when the channel between the two users is relatively
good on average.

There has also been some recent work on user cooperation systems
under various assumptions on the available channel state
information, and the level of cooperation among the users.
Laneman, Tse and Wornell \cite{laneman_tse_wornell} have
characterized the outage probability behavior for a system where
the users are allowed to cooperate only in half-duplex mode, and
 where no channel state information is available at the transmitters.
For the relay channel, which is a special one-sided case of user
cooperation, Host-Madsen and Zhang \cite{madsen} have solved for
power allocation policies that optimize some upper and lower
bounds on the ergodic capacity when perfect channel state
information is available at the transmitters and the receiver. For
a user cooperation system with finite capacity cooperation links,
Erkip \cite{erkip} has proposed a suboptimal solution to the
problem of maximizing the sum rate in the presence of full channel
state information, where it was also noted that the resulting
optimization problem is non-convex.

In this paper, we consider a two user fading cooperative Gaussian MAC
with complete channel state information at the transmitters and the
receiver, and average power constraints on the transmit powers. Note
that, this requires only a small quantity of additional feedback,
namely the amplitude information on the forward links, over the
systems requiring coherent combining \cite{coop_div}. In this case,
the transmitters can adapt their coding strategies as a function of
the channel states, by adjusting their transmit powers
\cite{goldsmith1, knopp_humblet, polymatroid}. We characterize the
optimal power allocation policies which maximize the set of ergodic
rates achievable by block Markov superposition coding. To this end, we
first prove that the seemingly non-concave optimization problem of
maximizing the achievable rates can be reduced to a concave problem,
by noting that some of the transmit power levels are essentially zero
at every channel state, which also reduces the dimensionality of the
problem. By this, we also show that the block Markov superposition
coding strategy proposed in \cite{generalized_feedback} and employed
in \cite{coop_div} for a Gaussian channel can be simplified
considerably by making use of the channel state information. Due to
the non-differentiable nature of the objective function, we use
sub-gradient methods to obtain the optimal power distributions that
maximize the achievable rates, and we provide the corresponding
achievable rate regions for various fading distributions. We
demonstrate that controlling the transmit powers in conjunction with
user cooperation provides significant gains over the existing rate
regions for cooperative systems.

\section{System Model}

\begin{figure}[!p]
\centerline{\epsfig{figure=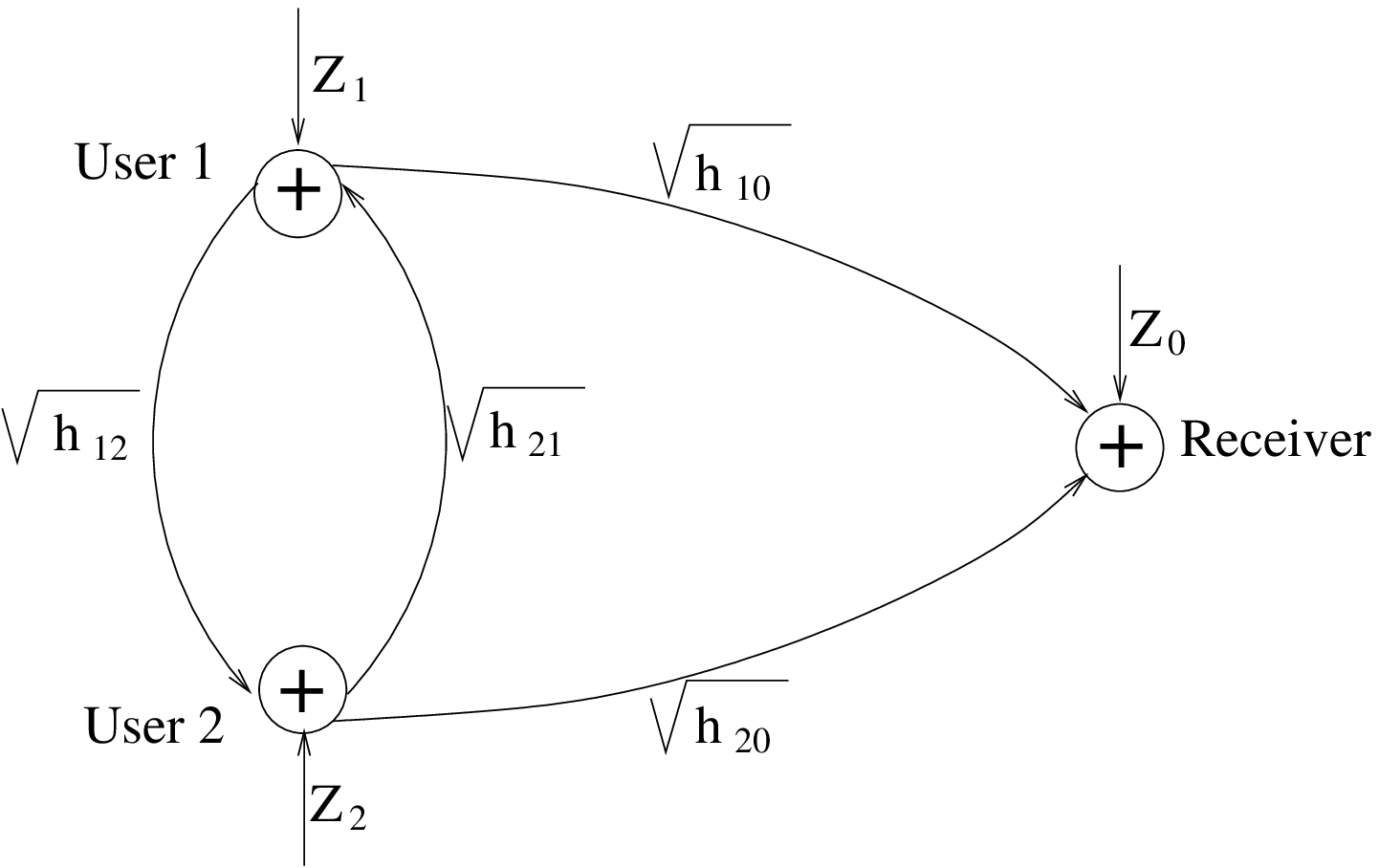, height=2.5in}}
\caption{Two user fading cooperative MAC.} \label{coopfig}
\end{figure}

We consider a two user fading Gaussian MAC, where both the
receiver and the transmitters receive noisy versions of the
transmitted messages, as illustrated in Figure~\ref{coopfig}. The
system is modelled by,
\begin{align} \label{model}
Y_0&=\sqrt{h_{10}}X_1+\sqrt{h_{20}}X_2+Z_0 \\
Y_1&=\sqrt{h_{21}}X_2+Z_1 \\
Y_2&=\sqrt{h_{12}}X_1+Z_2
\end{align}
where $X_i$ is the symbol transmitted by node $i$, $Y_i$ is the
symbol received at node $i$, and the receiver is denoted by $i=0$;
$Z_i$ is the zero-mean additive white Gaussian noise at node $i$,
having variance $\sigma_i^2$, and $\sqrt{h_{ij}}$ are the random
fading coefficients, the instantaneous realizations of which are
assumed to be known by both the transmitters and the receiver. We
assume that the channel variation is slow enough so that the
fading parameters can be tracked accurately at the transmitters,
yet fast enough to ensure that the long term ergodic properties of
the channel are observed within the blocks of transmission
\cite{biglieri}.

The transmitters are capable of making decoding decisions based on
the signals they receive and thus can form their transmitted
codewords not only based on their own information, but also based
on the information they have received from each other. This
channel model is a special case of the MAC with generalized
feedback \cite{generalized_feedback}. The achievable rate region
is obtained by using a superposition block Markov encoding scheme,
together with backward decoding, where the receiver waits to
receive all $B$ blocks of codewords before decoding. For the
Gaussian case, the superposition block Markov encoding is realized
as follows \cite{coop_div}: the transmitters allocate some of
their powers to establish some common information in every block,
and in the next block, they coherently combine part of their
transmitted codewords. In the presence of channel state
information, by suitably modifying the coding scheme given by
\cite{coop_div} to accommodate for channel adaptive coding
strategies, the encoding is performed by
\begin{equation}
X_i=\sqrt{p_{i0}(\hv)}X_{i0}+\sqrt{p_{ij}(\hv)}X_{ij}+\sqrt{p_{U_i}(\hv)}U_i
\end{equation}
for $\quad i,j~\in~\{1,2\}, \quad i\neq j$, where $X_{i0}$ carries
the fresh information intended for the receiver, $X_{ij}$ carries
the information intended for transmitter $j$ for cooperation in
the next block, and $U_i$ is the common information sent by both
transmitters for the resolution of the remaining uncertainty from
the previous block, all chosen from unit-power Gaussian
distributions. All the transmit power is therefore captured by the
power levels associated with each component, i.e., $p_{i0}(\hv)$,
$p_{ij}(\hv)$ and $p_{U_i}(\hv)$, which are required to satisfy
the average power constraints,
\begin{equation}\label{powercons}
E\left[p_{i0}(\hv)+p_{ij}(\hv)+p_{U_i}(\hv)\right]=E[p_i(\hv)]\leq
\pb_i, \quad i=1,2.
\end{equation}

Following the results in \cite{coop_div}, it can be shown that the
achievable rate region is given by the convex hull of all rate
pairs satisfying
\begin{eqnarray}
\label{rate1} R_{1}&<&E\left[\log\left(1+
\frac{h_{12}p_{12}(\hv)}{h_{12}p_{10}(\hv)+\sigma_2^2}\right)+\log\left(1+
\frac{h_{10}p_{10}(\hv)}{\sigma_0^2}\right)\right] \\
\label{rate2} R_{2}&<&E\left[\log\left(1+
\frac{h_{21}p_{21}(\hv)}{h_{21}p_{20}(\hv)+\sigma_1^2}\right)+\log\left(1+
\frac{h_{20}p_{20}(\hv)}{\sigma_0^2}\right)\right] \\
\label{ratesum} R_{1}+R_{2}&<&\min \left \{E\left[\log\left(1+
\frac{h_{10}p_{1}(\hv)+h_{20}p_{2}(\hv)+2\sqrt{h_{10}h_{20}p_{U_1}(\hv)p_{U_2}(\hv)}}{\sigma_0^2}\right)
\right], \right. \nonumber \\
 &&\left. \qquad \;\;E\left[\log\left(1+
\frac{h_{12}p_{12}(\hv)}{h_{12}p_{10}(\hv)+\sigma_2^2}\right)+\log\left(1+
\frac{h_{21}p_{21}(\hv)}{h_{21}p_{20}(\hv)+\sigma_1^2}\right)\right]
\right. \nonumber \\
&&\left. \qquad\;\;
+E\left[\log\left(1+\frac{h_{10}p_{10}(\hv)+h_{20}p_{20}(\hv)}{\sigma_0^2}\right)\right]\vphantom{\log\left(1+
\frac{h_{10}p_{1}(\hv)+h_{20}p_{2}(\hv)+2\sqrt{h_{10}h_{20}p_{U_1}(\hv)p_{U_2}(\hv)}}{\sigma_0^2}\right)}
\right\}
\end{eqnarray}
where the convex hull is taken over all power allocation policies
that satisfy (\ref{powercons}).

For a given power allocation, the rate region in
(\ref{rate1})-(\ref{ratesum}) is either a pentagon or a triangle,
since, unlike the traditional MAC, the sum rate constraint in
(\ref{ratesum}) may dominate the individual rate constraints
completely. The achievable rate region may alternatively be
represented as the convex hull of the union of all such regions.
Our goal is to find the power allocation policies that maximize
the rate tuples on the rate region boundary. \vspace{-15pt}

\section{Structure of the Sum Rate and the Optimal Policies}

We first consider the problem of optimizing the sum rate of the
system, as it will also shed some light onto the optimization of
an arbitrary point on the rate region boundary. The sum rate
(\ref{ratesum}) is not a concave function of the vector of
variables $\pv(\hv)=[p_{10}(\hv)~p_{12}(\hv)~ p_{U_1}(\hv)~
p_{20}(\hv) \\ ~p_{21}(\hv)~p_{U_2}(\hv)]$, due to the existence
of variables in the denominators. In what follows, we show that
for the sum rate to be maximized, for every given $\hv$, at least
two of the four components of  $[p_{10}(\hv)~ p_{12}(\hv)~
p_{20}(\hv)~ p_{21}(\hv)]$ should be equal to zero, which reduces
the dimensionality of the problem and yields a concave
optimization problem.

\begin{proposition}
Let the effective channel gains normalized by the noise powers be
defined as $s_{ij}=h_{ij}/\sigma_j^2$. Then, for the power control
policy $\pv^*(\hv)$ that maximizes (\ref{ratesum}), we need

\begin{enumerate}\setlength{\itemsep}{0pt}
\item $p_{10}^*(\hv)=p_{20}^*(\hv)=0$, if $s_{12}>s_{10}$ and
$s_{21}>s_{20}$ \item $p_{10}^*(\hv)=p_{21}^*(\hv)=0$, if
$s_{12}>s_{10}$ and $s_{21} \leq s_{20}$ \item
$p_{12}^*(\hv)=p_{20}^*(\hv)=0$, if $s_{12}\leq s_{10}$ and
$s_{21}>s_{20}$
\end{enumerate}
\hspace{2mm}$\left.
\begin{tabular}[h]{lll}
4. & \hspace{-2mm}$p_{12}^*(\hv)=p_{21}^*(\hv)=0$ \\
& \hspace{-2mm}\qquad \quad {\footnotesize OR} \\
& \hspace{-2mm}$p_{10}^*(\hv)=p_{21}^*(\hv)=0$ \\
& \hspace{-2mm}\qquad \quad {\footnotesize OR} \\
& \hspace{-2mm}$p_{12}^*(\hv)=p_{20}^*(\hv)=0$
\end{tabular}\right\} \mbox{if}\; s_{12}\leq
s_{10}\;\mbox{and}\;s_{21}\leq s_{20} \nonumber$

\end{proposition}
\vspace{2mm}
\begin{Proof}
To simplify the notation, let us drop the dependence of the powers
on the channel states, whenever such dependence is obvious from
the context. Let $p_i=p_{i0}+p_{ij}+p_{U_i}$ be the total power
allocated to a given channel state. Let us define
\setcounter{equation}{8}
\begin{align}
\label{A}
A&=1+s_{10}p_1+s_{20}p_2+2\sqrt{s_{10}s_{20}p_{U_1}p_{U_2}}
\\
B&=\frac{1+s_{10}p_{10}+s_{20}p_{20}}{(1+s_{12}p_{10})(1+s_{21}p_{20})}\\
C&=\left(1+s_{12}(p_{10}+p_{12})\right)\left(1+s_{21}(p_{20}+p_{21})\right)
\end{align}
Then, an equivalent representation of the sum rate in (\ref{ratesum})
is
\begin{equation}\label{sumcap}
R_{\s}=\min\left\{E\left[\log(A)\right],E\left[\log(BC)\right]\right\}
\end{equation}

Now, let us arbitrarily fix the total power level, $p_i$, as well
as the power level used for cooperation signals, $p_{U_i}\leq
p_i$, allocated to a given state for each user. For each such
allocation, the quantities $A$ and $C$ appearing in the sum rate
expression are fixed, i.e., allocating the remaining available
power $p_i-p_{U_i}$ among $p_{i0}$ and $p_{ij}$ will not alter
these quantities. Note that, such allocation also does not alter
the total power consumption at the given state, so we may limit
our attention to the maximization,
\begin{align}
\max_{\{p_{10},p_{20}\}} & \;\; B\left(p_{10},p_{20}\right) \nonumber \\
\mbox{s.t.} & \;\; p_{10}+p_{12} = p_1-p_{U_1} \nonumber \\
            & \;\;p_{20}+p_{21} = p_2-p_{U_2}
\end{align}
The partial derivatives of $B$ with respect to $p_{10}$ and
$p_{20}$ are
\begin{align}
\frac{\partial B}{\partial
p_{10}}=\frac{s_{10}-s_{12}(1+s_{20}p_{20})}{(1+s_{12}p_{10})^2(1+s_{21}p_{20})} \\
\frac{\partial B}{\partial
p_{20}}=\frac{s_{20}-s_{21}(1+s_{10}p_{10})}{(1+s_{21}p_{20})^2(1+s_{12}p_{10})}
\end{align}
Therefore, we make the following conclusions:
\begin{enumerate} \setlength\leftmargini  {1.3 em}
\item $s_{12}>s_{10}$, $s_{21}>s_{20}$. Then, $\frac{\partial
B}{\partial p_{10}}<0$ and $\frac{\partial B}{\partial p_{20}}<0$,
i.e., $B(p_{10},p_{20})$ is monotonically decreasing in both
$p_{10}$ and $p_{20}$, therefore the sum rate is maximized at
$p_{10}=p_{20}=0$.

\item $s_{12}>s_{10}$, $s_{21}\leq s_{20}$. Then, $\frac{\partial
B}{\partial p_{10}}<0$, and the function is maximized at
$p_{10}=0$ for any $p_{20}$. But this gives $\frac{\partial
B}{\partial p_{20}}|_{p_{10}=0}>0$, meaning $p_{20}$ should take
its maximum possible value, i.e., $p_{21}=0$.

\item $s_{12}\leq s_{10}$, $s_{21} > s_{20}$. Follows the same
lines of case 2 with roles of users 1 and 2 reversed.

\item $s_{12}\leq s_{10}$, $s_{21} \leq s_{20}$. In this case, the
partial derivatives of $B$ can be both made equal to zero within
the constraint set, yielding a critical point. However, using
higher order tests, it is possible to show that this solution
corresponds to a saddle point, and $B$ is again maximized at one
of the boundaries, $p_{10}=0$, $p_{20}=0$, $p_{10}=p_1-p_{U_1}$,
$p_{20}=p_2-p_{U_2}$. Inspection of the gradient on these boundary
points yields one of the three corner points
$\{(p_1-p_{U_1},0),(0,p_2-p_{U_2}),(p_1-p_{U_1},p_2-p_{U_2})\}$ as
candidates, each of which corresponds to one of the solutions in
case 4.

Although two of the components of the power vector are guaranteed to
be equal to zero, which ones will be zero depends on the $p_i$ and
$p_{U_i}$ that we fixed, therefore we are not able to completely
specify the solution, independent of $p_i$ and $p_{U_i}$, in this
case. On the other hand, the settings of interest to us are those
where the channels between the cooperating users are on average much
better than their direct links, since it is in these settings when
cooperative diversity yields high capacity gains \cite{coop_div}. In
such scenarios, the probability of both users' direct link gains
exceeding their corresponding cooperation link gains (case 4) is a
very low probability event. Therefore, which of the three possible
operating points is chosen is not of practical importance, and we can
safely fix the power allocation policy to one of them to carry on with
our optimization problem for the other variables. Although admittedly
this argument is likely to cause some suboptimality in our scheme, as
will be seen in the numerical examples, we still obtain a significant
gain in the achievable rates.
\end{enumerate}\end{Proof}

The significance of this result is two-fold. Firstly, given a
channel state, it greatly simplifies the well known block Markov
coding, in a very intuitive way: if the direct links of both users
are inferior to their cooperation links, the users do not transmit
direct messages to the receiver as a part of their codewords, and
they use each other as relays. If one of the users' direct channel
is better than its cooperation channel, and the other user is in
the opposite situation, then the user with the strong direct
channel chooses to transmit directly to the receiver, while the
weaker direct channel user still chooses to relay its information
over its partner. Second important implication of this result is
that it now makes the problem of solving for the optimal power
allocation policy more tractable, since it simplifies the
constraint set on the variables, and more importantly, this makes
the sum rate a concave function over the reduced set of
constraints and variables.

\begin{corollary}
The sum rate $R_{\s}$ given by (\ref{ratesum}), (\ref{sumcap}) is
a concave function of $\pv(\hv)$, over the reduced constraint set
described by Proposition 1.
\end{corollary}

\begin{Proof}
The proof of this result follows from directly substituting the
zero power components into the sum rate expression in
(\ref{sumcap}). Note that in each of the four cases, the second
function in the minimization, i.e., $\log(BC)$ takes either the
form $\log(1+a)+\log(1+b)$, or $\log(1+a+b)$, both of which are
clearly jointly concave in $a$ and $b$. Also, $\log(A)$ is clearly
a concave function of $\pv(\hv)$ since it is a composition of a
concave function with the concave and increasing logarithm. The
desired result is obtained by noting that the minimum of two
concave functions is concave.
\end{Proof}

Thus far we have discussed the structure of the sum rate, as well
as some properties of the optimal power allocation that maximizes
that rate. We now turn back to the problem of maximizing other
rate points on the rate region boundary. To this end, we point out
another remarkable property of the solution in Proposition 1.
Consider maximizing the bound on $R_1$ in (\ref{rate1}). For fixed
$p_{U_1}$ and $p_1$, it is easy to verify that all of the
available power should be allocated to the channel with the higher
gain, i.e., if $s_{12}>s_{10}$, then we need $p_{10}=0$ and
$p_{12}=p_1-p_{U_1}$. The same result also applies to $R_2$. But
this shows that, the policies described in Proposition 1
completely agree with optimal policies for maximizing the
individual rate constraints in cases 1-3, and they also agree if
we choose the operating point in case 4 to be
$p_{12}=0,\;p_{21}=0$. Therefore, the allocation in Proposition 1
enlarges the entire rate region in all directions (except for the
subtlety in case 4 for the sum rate). This has the benefit that
the weighted sum of rates, say $R_{\muv}=\mu_1R_1+\mu_2R_2$ also
has the same concavity properties of the sum rate, since for
$\mu_i>\mu_j$, the weighted sum of rates can be written as
$R_{\muv}=\mu_jR_{\s}+(\mu_i-\mu_j)R_i$, where both $R_{\s}$ and
$R_i$ are concave. Optimum power control policies that achieve the
points on the boundary of the achievable rate region can then be
obtained by maximizing the weighted sum of rates, which is the
goal of the next section.

\section{Rate Maximization via Subgradient Methods}

In this section we focus on maximizing the weighted sum of rates.
To illustrate both the results of the preceding section and the
problem statement for this section more precisely, let us
consider, without loss of generality, the case when $\mu_1 \geq
\mu_2$, and write down the optimization problem explicitly:
\begin{align} \label{weightedrates}
\max_{\pv(\hv)} \;&
(\mu_1-\mu_2)\left\{\vphantom{\int}E_{1,2}\left[\log(1+p_{12}(\hv)s_{12})\right]\right.
\nonumber
\\
& +\left.\vphantom{\int}
E_{3,4}\left[\log(1+p_{10}(\hv)s_{10})\right]\right\} +
\mu_2 \min \left\{\vphantom{\int}E[\log(A)],\right. \nonumber \\
& +
E_1\left[\log(1+p_{12}(\hv)s_{12})+\log(1+p_{21}(\hv)s_{21})\right]
 \nonumber \\
& +
E_2\left[\log(1+p_{12}(\hv)s_{12})+\log(1+p_{20}(\hv)s_{20})\right]
\nonumber \\
& +
E_3\left[\log(1+p_{10}(\hv)s_{10})+\log(1+p_{21}(\hv)s_{21})\right]
 \nonumber \\
& + \left.
E_4\left[\log(1+p_{10}(\hv)s_{10}+p_{20}(\hv)s_{20})\right]
\vphantom{\int}\right\} \nonumber \\
\mbox{s.t.} \;&
E_{3,4}\left[p_{10}(\hv)\right]+E_{1,2}\left[p_{12}(\hv)\right]+E\left[p_{U_1}(\hv)\right]\leq
\pb_1 \nonumber \\&
E_{2,4}\left[p_{20}(\hv)\right]+E_{1,3}\left[p_{21}(\hv)\right]+E\left[p_{U_2}(\hv)\right]\leq
\pb_2
\end{align}
where, $E_S$ denotes the expectation over the event that case $S
\subset \{1,2,3,4\}$  from Proposition 1 occurs, and $A$ is as
given by (\ref{A}). Note that the objective function is concave,
and the constraint set is convex, therefore we conclude that any
local optimum for the constrained optimization problem is a global
optimum. Although $R_{\muv}$ is differentiable almost everywhere
since it is concave, its optimal value is attained along the
discontinuity of its gradient, namely when the two arguments of
the min operation are equal. Hence, we solve the optimization
problem using the method of subgradients from non-differentiable
optimization theory \cite{shor,bertsekasNP}.

The subgradient methods are very similar to gradient ascent
methods in that whenever the function is differentiable (in our
case almost everywhere), the subgradient is equal to the gradient.
However, their major difference from gradient ascent methods is
that they are not necessarily monotonically non-decreasing. Let us
denote the objective function in (\ref{weightedrates}) by
$R_{\muv}$. A subgradient for the concave function $R_{\muv}(\pv)$
is any vector $\gv$ that satisfies \cite{bertsekasNP},

\begin{equation}
R_{\muv}(\pv')\leq R_{\muv}(\pv)+ (\pv'-\pv)^{\top}\gv
\end{equation}
and the subgradient method for the constrained maximization uses the
update \cite{bertsekasNP}

\begin{equation}
\pv(k+1)=[\pv(k)+\alpha_k\gv_k]^{\dag}
\end{equation}
where $[\cdot]^{\dag}$ denotes the Euclidean projection onto the
constraint set, and $\alpha_k$ is the step size at iteration $k$.
There are various ways to choose $\alpha_k$ to guarantee
convergence to the global optimum; for our particular problem, we
choose the diminishing stepsize, normalized by the norm of the
subgradient to ensure the convergence \cite{shor}
\begin{equation}
\alpha_k=\frac{a}{b+\sqrt{k}}\frac{1}{\parallel \gv \parallel}
\end{equation}

While the subgradient method is a powerful tool to obtain the
global optimum of (\ref{weightedrates}), it does not shed light
onto the characteristics of the optimal power allocation policies.
Although it is in general difficult, if not impossible, to state
the optimal policy in a closed form, we provide a simplified case
as an example, and provide conditions the optimal policies should
satisfy, in the next subsection.

\subsection{Properties of the Optimal Power Allocation: An Example}

Throughout the analysis in this subsection, we assume for simplicity
of the derivations that the fading distributions are such that all
realizations of the fading values satisfy $s_{12}>s_{10}$ and
$s_{21}>s_{20}$, i.e., the system always operates in the more
interesting case 1. This particular case is of practical interest
since the cooperating transmitters are likely to be closely located
with less number of scatterers and obstructions when compared to their
paths to the receiver, and thus have better channel conditions among
each other. We also consider maximizing the sum capacity, i.e.,
$\mu_1=\mu_2$. Under these assumptions, since
$p_{10}(\hv)=p_{20}(\hv)=0$ always, the problem of maximizing the sum
rate, given in (\ref{ratesum}) and also obtained from
(\ref{weightedrates}) by inserting $\mu_1=\mu_2$, reduces to
\begin{align}
\max_{\pv(\hv)} \;\;& \min \left \{E\left[\log(D) \right],
E\left[\log\left(1+s_{12}p_{12}(\hv)\right)+\log\left(1+s_{21}p_{21}(\hv)\right)\right]\right\}
\nonumber \\
\mbox{s.t.} \;\;& E\left[p_{12}(\hv)+p_{U_1}(\hv)\right]\leq \pb_1 \nonumber\\
& E\left[p_{21}(\hv)+p_{U_2}(\hv)\right]\leq \pb_2 \nonumber \\
& p_{12}(\hv),p_{U_1}(\hv),p_{21}(\hv),p_{U_2}(\hv) \geq 0, \quad
\forall \hv
\end{align}
where
\begin{equation}\label{D}
D=1+s_{10}\left(p_{12}(\hv)+p_{U_1}(\hv)\right)+s_{20}\left(p_{21}(\hv)+p_{U_2}(\hv)\right)+2\sqrt{s_{10}s_{20}p_{U_1}(\hv)p_{U_2}(\hv)}
\end{equation}

Note that, the maximum of the objective function is attained only when
the arguments of the $\min$ operation are equal, or else it would be
possible to increase the smaller argument while decreasing the larger,
just by transferring the transmit power between the cooperation
signals $p_{U_i}(\hv)$ and $p_{ij}(\hv)$ at some $\hv$, while keeping
their sum constant, and therefore not violating the average power
constraints. Hence, to overcome the difficulties arising from the
non-differentiability of the cost function, the $\min$ in the cost
function can be replaced by an equality constraint in the constraint
set, and the problem can be restated with a new differentiable
objective function as
\begin{align}
\max_{\pv(\hv)} \;\;&
E\left[\log\left(1+s_{12}p_{12}(\hv)\right)+\log\left(1+s_{21}p_{21}(\hv)\right)\right] \label{examplestatement1} \\
\mbox{s.t.} \;\;& E\left[\log(D) \right]=
E\left[\log\left(1+s_{12}p_{12}(\hv)\right)+\log\left(1+s_{21}p_{21}(\hv)\right)\right] \label{examplestatement2} \\
& E\left[p_{12}(\hv)+p_{U_1}(\hv)\right]\leq \pb_1 \label{examplestatement3} \\
& E\left[p_{21}(\hv)+p_{U_2}(\hv)\right]\leq \pb_2 \label{examplestatement4} \\
& p_{12}(\hv),p_{U_1}(\hv),p_{21}(\hv),p_{U_2}(\hv) \geq 0, \quad
\forall \hv \label{examplestatement5}
\end{align}

Associating the Lagrange multiplier $\gamma$ to the equality
constraint (\ref{examplestatement2}), $\lambda_1,\lambda_2>0$ to
the power constraints
(\ref{examplestatement3}),(\ref{examplestatement4}), and
$\xi_i(\hv)\geq 0$, $i=1,\cdots,4$ to the non-negativity
constraints (\ref{examplestatement5}), and noting that the power
constraints need to be satisfied by equality, we obtain the
Karush-Kuhn-Tucker (KKT) necessary conditions for optimality. Note
that for optimality, for any given $\hv$, the components
$p_{U_1}(\hv)$ and $p_{U_2}(\hv)$ should be either both positive
or both zero. Therefore, it is sufficient to analyze these two
cases.

First let $p_{U_1}(\hv),p_{U_2}(\hv)>0$. Then, the KKT conditions
reduce to
\begin{align}
(1+\gamma)\frac{s_{12}}{1+s_{12}p_{12}(\hv)}&\leq\lambda_1+\gamma\frac{s_{10}}{D} \label{KKT1}\\
(1+\gamma)\frac{s_{21}}{1+s_{21}p_{21}(\hv)}&\leq\lambda_2+\gamma\frac{s_{20}}{D} \label{KKT2}\\
\gamma
\frac{\sqrt{s_{10}s_{20}p_{U_2}(\hv)}+s_{10}\sqrt{p_{U_1}(\hv)}}{D\sqrt{p_{U_1}(\hv)}}
&= -\lambda_1 \label{KKT3}\\
\gamma
\frac{\sqrt{s_{10}s_{20}p_{U_1}(\hv)}+s_{20}\sqrt{p_{U_2}(\hv)}}{D\sqrt{p_{U_2}(\hv)}}
&= -\lambda_2 \label{KKT4}
\end{align}
From (\ref{KKT3}) and (\ref{KKT4}), it follows that there is a
linear relationship between the optimal $p_{U_1}(\hv)$ and
$p_{U_2}(\hv)$, i.e.,
\begin{equation} \label{linear}
\lambda_1^2s_{20}p_{U_1}(\hv)=\lambda_2^2s_{10}p_{U_2}(\hv)
\end{equation}

Next, we include the case when $p_{U_1}(\hv)=p_{U_2}(\hv)=0$ in
our analysis. The partial derivatives $\frac{\partial D}{\partial
p_{U_i}(\hv)}$, $i=1,2$ are not well defined when $p_{U_i}(\hv)$
are both zero. However, the above derivation shows that the
optimal solution is guaranteed to lie on the line given by
(\ref{linear}), since $(p_{U_1}(\hv),p_{U_2}(\hv))=(0,0)$ also
lies on the same line. Therefore, it is sufficient to evaluate the
optimality conditions along the line (\ref{linear}) for
$p_{U_1}(\hv)=p_{U_2}(\hv)=0$, as well as for
$p_{U_1}(\hv),p_{U_2}(\hv)>0$. Substituting (\ref{KKT3}) into
(\ref{KKT1}), (\ref{KKT4}) into (\ref{KKT2}), using (\ref{linear}),
and extending (\ref{KKT3}) and (\ref{KKT4}) to include the case
$p_{U_1}(\hv),p_{U_2}(\hv)=(0,0)$, we obtain the overall set of
necessary conditions for the optimality of the power allocation
policy as,
\begin{align}
\frac{s_{12}}{1+s_{12}p_{12}(\hv)}&\leq\frac{1}{(1+\gamma)}\frac{\lambda_1^2s_{20}}{\lambda_2s_{10}+\lambda_1s_{20}} &\mbox{w.e. if}\; p_{12}(\hv)>0 \label{KKTfinal1}\\
\frac{s_{21}}{1+s_{21}p_{21}(\hv)}&\leq\frac{1}{(1+\gamma)}\frac{\lambda_2^2s_{10}}{\lambda_2s_{10}+\lambda_1s_{20}} &\mbox{w.e. if}\; p_{21}(\hv)>0 \label{KKTfinal2}\\
-\lambda_1& \leq \frac{\gamma}{D}\left(s_{10}+\frac{\lambda_1}{\lambda_2}s_{20}\right) &\mbox{w.e. if}\; p_{U_1}(\hv)>0 \label{KKTfinal3}\\
-\lambda_2& \leq
\frac{\gamma}{D}\left(s_{20}+\frac{\lambda_2}{\lambda_1}s_{10}\right)
&\mbox{w.e. if}\; p_{U_2}(\hv)>0\label{KKTfinal4}
\end{align}
where w.e. stands for ``with equality.''

These conditions shed some light onto the structure of the optimal
power allocation policy: when the channel gains $s_{10},s_{20}$
from the transmitters to the receiver are fixed, the optimal power
levels associated with the components intended for the other
transmitter, $p_{12}(\hv)$ and $p_{21}(\hv)$, are independent of
each other and are obtained by {\em single user waterfilling} over
the individual link gains $s_{12}$ and $s_{21}$, respectively. The
direct link gains $s_{10}$ and $s_{20}$ determine the water level.
This is easily observed by denoting the right hand sides of
(\ref{KKTfinal1}) and (\ref{KKTfinal2}) by the constants $\nu_1$
and $\nu_2$ (since $s_{10}$ and $s_{20}$ are fixed), and solving
for $p_{12}(\hv)$ and $p_{21}(\hv)$, keeping in mind the
equalities are satisfied only when the power levels are positive,
i.e.,
\begin{align}
p_{12}(\hv)=\left(\frac{1}{\nu_1}-\frac{1}{s_{12}}\right)^+\\
p_{21}(\hv)=\left(\frac{1}{\nu_2}-\frac{1}{s_{21}}\right)^+
\end{align}
where $(\cdot)^+$ denotes $\max(0,\cdot)$.

Once $p_{12}(\hv)$ and $p_{21}(\hv)$ are calculated, they can be
substituted into (\ref{KKTfinal3}) and (\ref{KKTfinal4}) with $D$
defined as in (\ref{D}) to obtain $p_{U_i}$. On the other hand,
the computation of the optimal power levels require solving for
the Lagrange multipliers $\lambda_i$ and $\gamma$ that satisfy the
power constraints, which would require a multi-dimensional search.
Moreover, due to the constraint (\ref{KKT2}) being potentially
non-convex for some power values, the solution to the KKT
conditions may not be sufficient to guarantee optimality. In the
next section, we obtain the optimal power allocation policy
numerically using the subgradient method, and then demonstrate by
simulations that the optimal power allocation policy possesses the
properties described in this subsection.

\section{Simulation Results}
In this section we provide some numerical examples to illustrate
the performance of the proposed joint power allocation and
cooperation scheme.

\begin{figure}[!p]
\centerline{\epsfig{figure=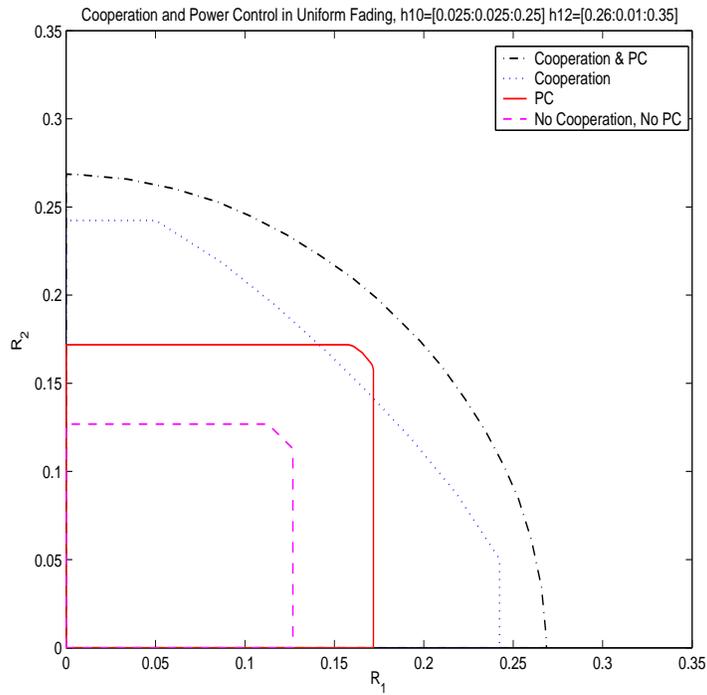,
height=3.7in,width=3.7in}} \caption{Rates achievable by joint
power control and user cooperation for uniform fading.}
\label{unifig}
\end{figure}

Figure \ref{unifig} illustrates the achievable rate region we
obtain for a system with $\pb_i=\sigma_i^2=1$, subject to uniform
fading, where the links from the transmitters to the receiver are
symmetric and take values from the set
$\{0.025,0.050,\cdots,0.25\}$, each with probability $1/10$, while
the link among the transmitters is also symmetric and uniform, and
takes the values $\{0.26,0.27,\cdots,0.35\}$. Notice that here, we
have intentionally chosen the fading coefficients such that the
cooperation link is always better than the direct links,
therefore, the system operates only in case 1 of Proposition 1.
Consequently, in this particular case, our power allocation scheme
is actually the optimal power allocation policy for the block
Markov superposition encoding scheme.

The region for joint power control and cooperation is generated using
the subgradient method with parameters $a=50$ and $b=5$. We carried
out the optimization for various values of the priorities $\mu_i$ of
the users, each of which give a point on the rate region boundary, and
then we performed a convex hull operation over these points. We
observe that power control \cite{goldsmith1, knopp_humblet,
polymatroid} by itself improves on the rate region of the cooperative
system with no power control, for rate pairs close to the sum rate, by
utilizing the direct link more efficiently. Joint user cooperation and
power control scheme significantly improves on all other schemes, as
it takes advantage of both cooperation diversity and time diversity in
the system. In fact, we can view this joint diversity utilization as
adaptively performing coding, medium accessing and routing, thereby
yielding a cross-layer approach for the design of the wireless
communication system.

\begin{figure}[!p]
\centerline{\epsfig{figure=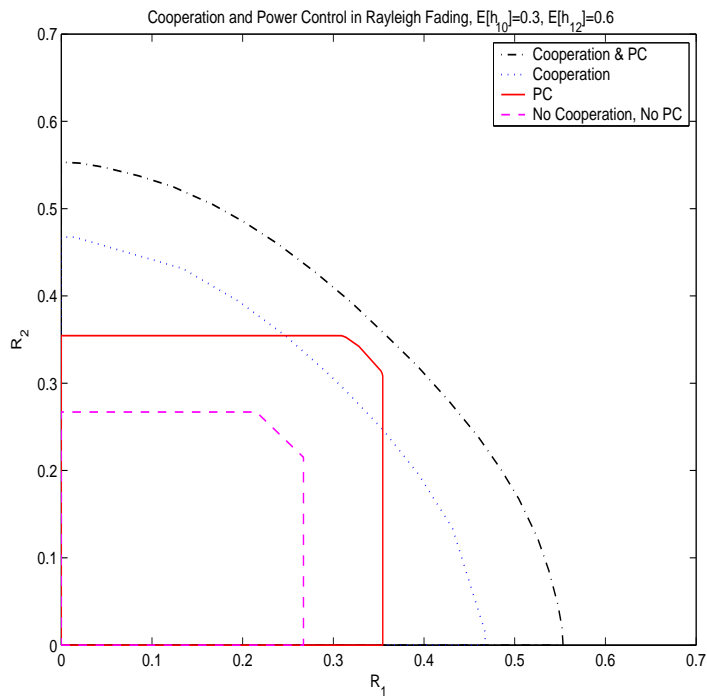,
height=3.7in,width=3.7in}} \caption{Rates achievable by joint
power control and user cooperation for Rayleigh fading.}
\label{rayfig}
\end{figure}

Figure \ref{rayfig} also corresponds to a system with unit SNR,
but this time subject to Rayleigh fading, i.e., the power gains to
the receivers are exponential random variables, with
$E[h_{10}]=E[h_{20}]=0.3$, $E[h_{12}]=E[h_{21}]=0.6$. In this
setting, all four cases in Proposition 1 are realized, and there
is potentially some loss over the optimally achievable rates.
However, we obtain a very similar set of rate regions to the
uniform case, indicating that in fact the loss, if any, is very
small thanks to the very low probability of both of the direct
links outperforming the cooperation link.

It is interesting to note in both Figures~\ref{unifig}
and~\ref{rayfig}, cooperation with power control improves
relatively less over power control only near the sum capacity.
This can be attributed to the fact that, for the traditional MAC,
the sum rate is achieved by time division among the users, which
does not allow for coherent combining gain \cite{knopp_humblet}.
Therefore, it is not surprising to see that in order to attain
cooperative diversity gain, users may have to sacrifice some of
the gain they obtain from exploiting the time diversity.

\begin{figure}[!t]
\centerline{\epsfig{figure=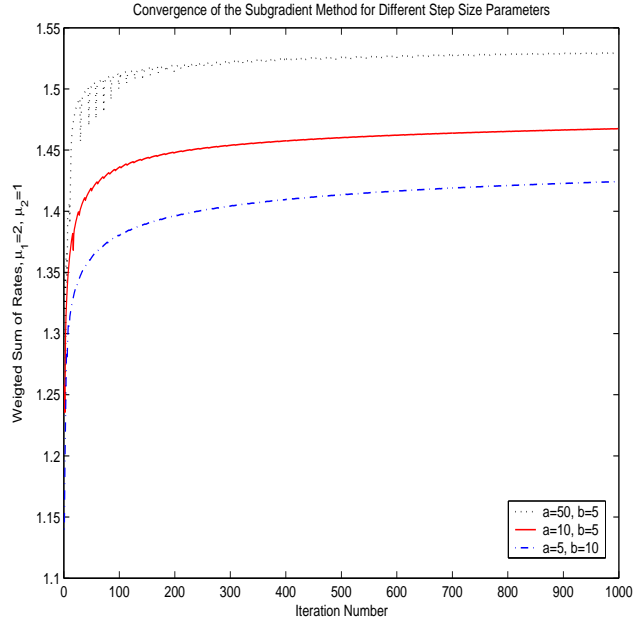,
height=3.3in,width=3.3in}} \caption{Convergence of $R_{\muv}$
using subgradient method for different step size parameters.}
\label{convfig}
\end{figure}

In Figure~\ref{convfig}, we illustrate the convergence of the
subgradient method. The objective function is $R_{\muv}$ with
$\mu_1=2$ and $\mu_2=1$, and the step size parameters are varied.
We observe that, by choosing larger step sizes, the non-monotonic
behavior of the subgradient algorithm becomes more apparent,
however the convergence is significantly faster than the smaller
step sizes, as the algorithm is more likely to get near the
optimal value of the function in the initial iterations. Note
that, in our simulations we terminated the algorithm after $1000$
iterations, and the three curves would eventually converge after
sufficiently large number of iterations.

\begin{figure}[p]
\centering \subfigure[hang][The power level $p_{12}(\hv)$ as a
function of $s_{12}$ and $s_{21}$.]{\label{cooppcjournal1}
\epsfig{figure=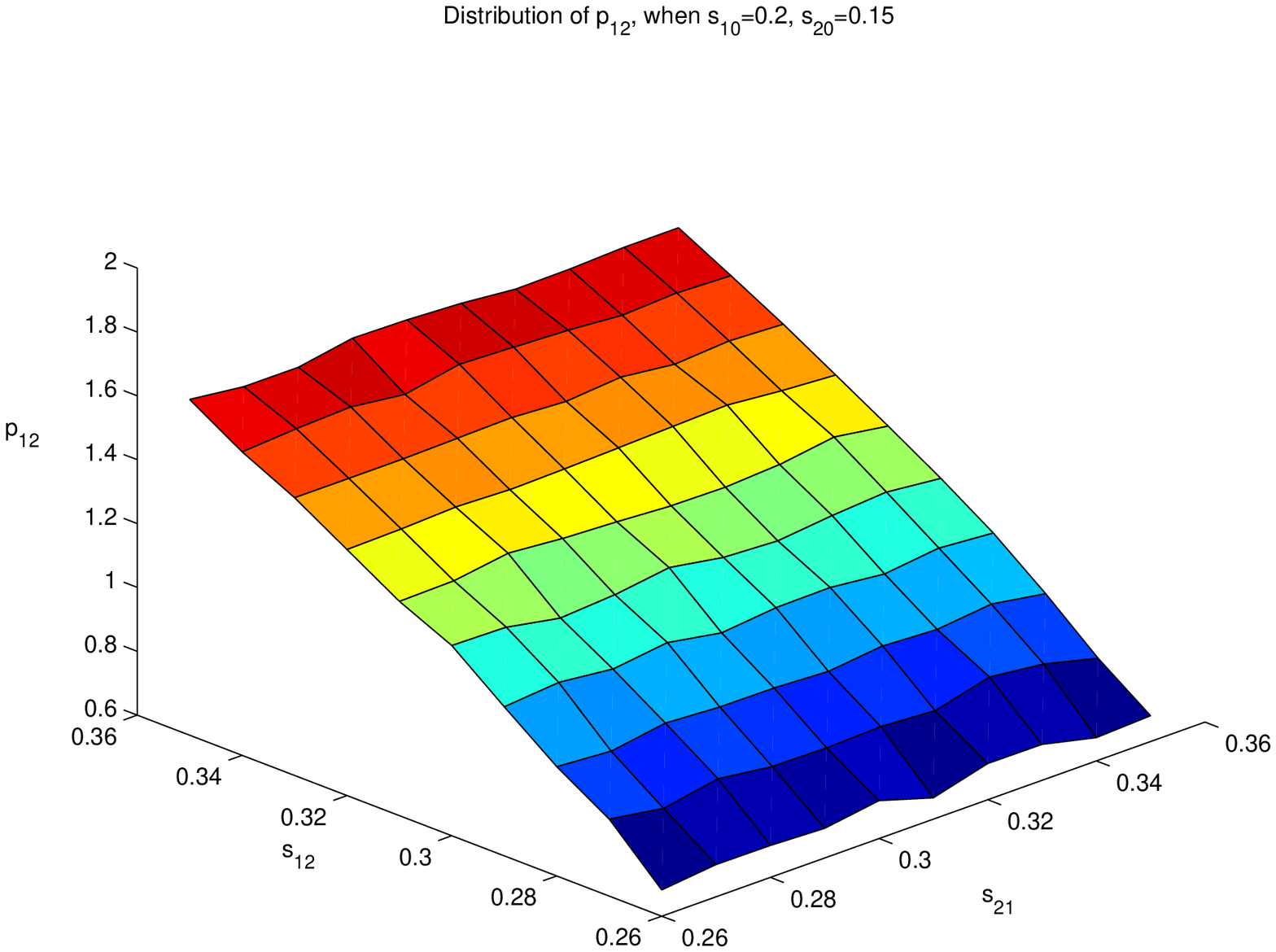, height=3in,width=3in}}
\subfigure[hang][The power level $p_{21}(\hv)$ as a function of
$s_{12}$ and $s_{21}$.]{\label{cooppcjournal2}
\epsfig{figure=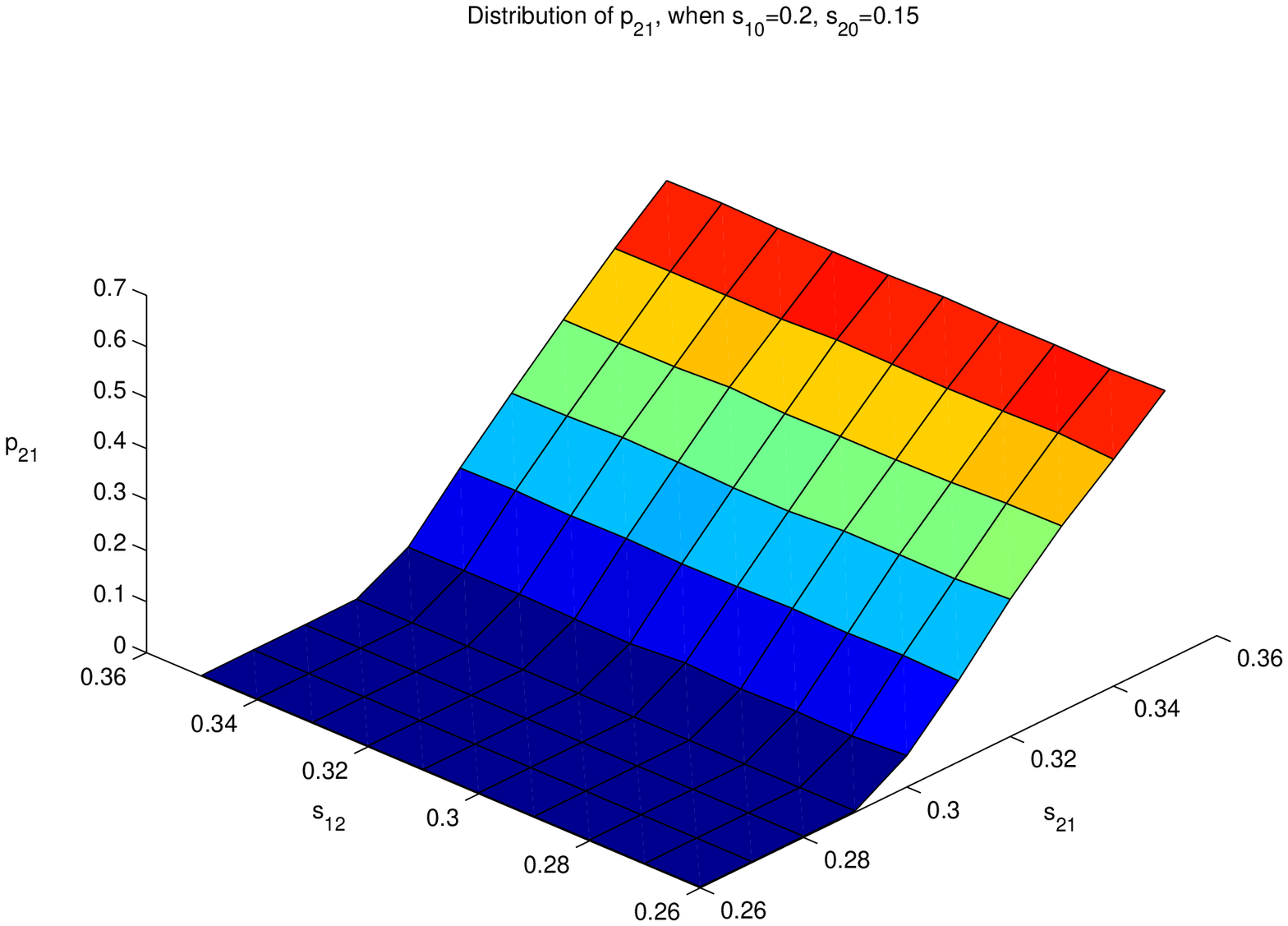, height=3in,width=3in}}
\caption{Power distributions obtained using the projected
subgradient method.}
\end{figure}

Figures~\ref{cooppcjournal1} and \ref{cooppcjournal2} demonstrate
the waterfilling nature of the optimal power levels $p_{12}(\hv)$,
and $p_{21}(\hv)$, as obtained in Section 4.1. These power levels
correspond to the sum rate point on Figure~\ref{unifig}, and were
obtained by using the subgradient algorithm, after 1000
iterations. The direct link gains are fixed to the values
$s_{10}=0.2$ and $s_{20}=0.15$. As dictated by (\ref{KKTfinal1})
and (\ref{KKTfinal2}), the ``water level'' for $p_{12}$ is higher,
and the threshold channel state level to start transmitting is
lower. We see that after 1000 iterations, the subgradient approach
nearly satisfies the optimality conditions.

\section{Conclusions}

We have addressed the problem of optimal power allocation for a
fading cooperative MAC, where the transmitters and the receiver
have channel state information, and are therefore able to adapt
their coding and decoding strategies by allocating their
resources. We have characterized the power control policies that
maximize the rates achievable by block Markov superposition
coding, and proved that, in the presence of channel state
information, the coding strategy is significantly simplified:
given any channel state, for each of the users, among the three
signal components, i.e., those that are intended for the receiver,
for the other transmitter, and for cooperation, at least one of
the first two should be allocated zero power at that channel
state. This result also enabled us to formulate the otherwise
non-concave problem of maximizing the achievable rates as a
concave maximization problem. The power control policies, which
are jointly optimal with block Markov coding, were then obtained
using subgradient method for non-differentiable optimization. The
resulting achievable rate regions for joint power control and
cooperation improve significantly on cooperative systems without
power control, since our joint approach makes use of both
cooperative diversity and time diversity.

\clearpage

\end{document}